\documentclass[conference]{IEEEtran}

\usepackage{pgfplots}
\usepackage[binary-units]{siunitx}
\usepackage{color}
\usepackage{mathtools}
\usepackage[noend]{algpseudocode}
\usepackage{algorithm}
\usepackage{multirow}
\usepackage{booktabs}
\usepackage[hidelinks]{hyperref}

\begin{document}

\title{Stochastic Rounding: Algorithms and Hardware Accelerator}

\author{
\IEEEauthorblockN{
Mantas Mikaitis}
\IEEEauthorblockA{
Department of Mathematics \\
The University of Manchester \\
Email: mantas.mikaitis@manchester.ac.uk}}

\maketitle

\begin{abstract}
Algorithms and a hardware accelerator for performing stochastic rounding (SR) are presented.
The main goal is to augment the ARM M4F based multi-core processor SpiNNaker2 with a more flexible rounding functionality than is available in the ARM processor itself.
The motivation of adding such an accelerator in hardware is based on our previous results showing improvements in numerical accuracy of ODE solvers in fixed-point arithmetic with SR, compared to standard round to nearest or bit truncation rounding modes.
Furthermore, performing SR purely in software can be expensive, due to requirement of a pseudorandom number generator (PRNG), multiple masking and shifting instructions, and an addition operation.
Also, saturation of the rounded values is included, since rounding is usually followed by saturation, which is especially important in fixed-point arithmetic due to a narrow dynamic range of representable values.
The main intended use of the accelerator is to round fixed-point multiplier outputs, which are returned unrounded by the ARM processor in a wider fixed-point format than the arguments.
\end{abstract}

\begin{IEEEkeywords}
stochastic rounding, fixed-point arithmetic, floating-point arithmetic, bfloat16.
\end{IEEEkeywords}

\IEEEdisplaynontitleabstractindextext

\section{Introduction}

SpiNNaker is a 18-ARM968-core (integer only) chip for simulating neural networks, including ordinary differential equations (ODEs) of neurons \cite{spinnproject}.
Previous work on SpiNNaker \cite{doi:10.1098/rsta.2019.0052} explored numerical accuracy issues in ODE solvers run in fixed-point arithmetic with the main conclusion that rounding errors are a major factor in divergence of the solution from the reference double precision solution, and that stochastic rounding (SR) helps in reducing divergence.
It was also shown that fixed point with SR is $2.6-4.2\times$ faster than emulating floating-point arithmetic in software.

The next generation SpiNNaker, SpiNNaker2 will be based on an ARM Cortex-M4F processor \cite{hoppnerspinnaker2}, which does not have a capability of rounding a fixed-point number to a specified number of bits.
There are three instructions with rounding available: \texttt{SMMLAR} --- multiply two numbers, add a third number to the top 32 bits of the result and return the rounded 32 top bits; \texttt{SMMLSR} --- the same as \texttt{SMMLAR}, but subtract the third argument; and \texttt{SMMULR} --- multiply and return the rounded 32 top bits of the result \cite{arm_tech_reference}.
Rounding is done by adding 0x80000000 to the product, therefore the tie-breaking rule is round up \cite{arm_m4_user_guide}.
While this would work well for s16.15 $\times$ u0.32 multiplications (where $\{s/u\}X.Y$ is a signed/unsigned 2's complement fixed-point format with $X$ integer bits and $Y$ fractional bits), it is limited in terms of other mixed-format multiplications demonstrated in \cite{doi:10.1098/rsta.2019.0052}.
To implement round to nearest (RN) and stochastic rounding would require multiple instructions, usually working on two registers containing a 64-bit unrounded value.
Furthermore, there is no mention as to whether there are instructions available on the Cortex-M4F processor to perform \textit{saturation} after rounding (return a maximum representable value on overflow).
While saturation instructions for 32-bit values, with configurable saturation bit position and saturated addition, are available on the M4F, saturating a 64-bit value from the multiplication would need to be done by comparison and because it is a value across two registers, multiple instructions would be required to obtain a rounded and saturated value somewhere in the middle of a 64-bit value.
Additionally, since ARM M4F has a single-precision floating-point (binary32) \cite{ieee19} unit, it is beneficial to add binary32 to bfloat16 (equivalent to binary32 with the bottom 16 bits removed --- 1 sign bit, 8 exponent bits and 7 significand bits) \cite{intel_bfloat16} rounding, which is an elegant format for storage and can be operated on using binary32 hardware.

The contributions of this paper are as follows.
\begin{itemize}
	\item Two bit-level algorithms for doing stochastic rounding and saturation (Section~\ref{sec:algorithms}).
	\item The architecture of the accelerator for doing rounding and saturation (Sections~\ref{sec:specification}~and~\ref{sec:design}).
	\item Three accelerators with 8/16/32-bit random number precisions in stochastic rounding are evaluated in 22nm technology. Leakage and area comparisons are demonstrated (Section~\ref{sec:evaluation}).
\end{itemize}

\section{Algorithms}
\label{sec:algorithms}

Stochastic rounding has recently been explored in machine learning due to substantial improvements in reducing rounding errors in low precision numerical formats \cite{HOHFELD1992291,Gupta:2015:DLL:3045118.3045303,Wang2018, mellempudi2019mixed}.
The first mention of it can be traced back to \cite{doi:10.1137/1001011}.

Stochastic rounding differs from the standard rounding modes, such as round to nearest, in that instead of always rounding to the nearest number, the decision about which way to round is non-deterministic and the probability of rounding up is proportional to the \textit{residual} (value of the trailing bits that do not fit into the destination format interpreted to be in the range $[0, 1)$).
Given a real number $x$, an output fixed-point format to round the value to, $<s, i, p>$ (where \textit{s} tells us whether it is signed or unsigned format, \textit{i} defines the number of integer bits and \textit{p} defines the number of fractional bits); defining $\lfloor x \rfloor$ as the truncation operation (cancelling a number of bottom bits with values smaller than $\varepsilon=2^{-p}$ and leaving $p$ fractional bits) which returns a number in $<s, i, p>$ format less than or equal to $x$; and given a random value $P \in [0, 1)$, drawn from a uniform random number generator, SR is defined as
\begin{align}
\begin{split}
\mathrm{SR}(x, <s,i,p>) =
\left\{
\begin{matrix*}[l]
\lfloor x \rfloor & \text{if} \: P \geq \frac{x-\lfloor x \rfloor}{\varepsilon\mathstrut}, \\
\lfloor x \rfloor + \varepsilon & \text{if} \: P < \frac{x-\lfloor x \rfloor}{\varepsilon}. \\
\end{matrix*}
\right.
\label{equ:sr}
\end{split}
\end{align}
For floating-point numbers the definition of SR is slightly different since it does not use 2's complement, see for example \cite{Wang2018}.

\begin{algorithm}[b!]
	\caption{Stochastic rounding by comparison}
	\label{alg:stochastic_rounding_by_comparison}
	\begin{algorithmic}
		\Function{SATSR\_int64\_int32}{$X, n$}
		\State ${P} \gets \Call{PRNG32()}{}$
		\State ${\mathit{MASK}} \gets ((1\ll n)-1)$
		\State ${P} \gets P \& \mathit{MASK}$
		\State ${\mathit{RESIDUAL}} \gets X \& \mathit{MASK}$
		\State ${X} \gets X \gg n$
		\If {${P < \mathit{RESIDUAL}}$}
		\State ${X} \gets X+1$
		\EndIf
		\If {$X > \mathit{MAX\_INT32}$}
		\State return $\mathit{MAX\_INT32}$
		\EndIf
		\If {$X < \mathit{MIN\_INT32}$}
		\State return $\mathit{MIN\_INT32}$
		\EndIf
		\State return $X$
		\EndFunction
	\end{algorithmic}
\end{algorithm}

This can be implemented by inspecting residuals and utilizing a pseudorandom number generator (PRNG); in \cite{doi:10.1098/rsta.2019.0052} a simple \textit{linear congruential generator} as well as more complex ones were used without any significant differences in numerical results.
For the current work, SpiNNaker2 already has a generator from the family of generators called \textit{KISS} (proposed by George Marsaglia) implemented in hardware with configurable seeds \cite{hoppnerspinnaker2}, from which we fetch 32-bit random bit streams for rounding.
SpiNNaker2 hardware implementation is after \cite[p.~3]{Jone10} (algorithm called \texttt{JKISS32}) except one of the internal variables was modified to be 64 bits to improve the quality of the generator.

For a fully configurable stochastic rounding routine we have to be able to round a specified number of trailing bits $n$ of a 64-bit number.
This means that bits $n-1$ to $0$ are zeroed and 0x1 (hex value) is added at the $n$-th bit location if rounding is performed.
Usually there is no need to return the rounded value in the original bit width with bottom bits zeroed, therefore an output number from the rounding routine is provided in lower precision, at which point it also has to be saturated if the value is too large to be represented.
For SpiNNaker use cases we are interested in rounding a 64-bit multiplication result to various 32-bit fixed-point formats, therefore we explored an algorithm and implementation for this routine.

\begin{algorithm}
	\caption{Stochastic rounding by addition}
	\label{alg:stochastic_rounding_by_addition}
	\begin{algorithmic}
		\Function{SATSR\_int64\_int32}{$X, n$}
		\State ${P} \gets \Call{PRNG32()}{}$
		\State ${P} \gets P \& ((1\ll n)-1)$
		\State ${X} \gets (X+P) \gg n$
		\If {$X > \mathit{MAX\_INT32}$}
		\State return $\mathit{MAX\_INT32}$
		\EndIf
		\If {$X < \mathit{MIN\_INT32}$}
		\State return $\mathit{MIN\_INT32}$
		\EndIf
		\State return $X$
		\EndFunction
	\end{algorithmic}
\end{algorithm}

\begin{algorithm}
	\caption{Rounding to nearest with round up on a tie}
	\label{alg:rounding_to_nearest}
	\begin{algorithmic}
		\Function{SATRN\_int64\_int32}{$X, n$}
		\State ${X} \gets (X+(1\ll (n-1))) \gg n$
		\If {$X > \mathit{MAX\_INT32}$}
		\State return $\mathit{MAX\_INT32}$
		\EndIf
		\If {$X < \mathit{MIN\_INT32}$}
		\State return $\mathit{MIN\_INT32}$
		\EndIf
		\State return $X$
		\EndFunction
	\end{algorithmic}
\end{algorithm}

There are two simple ways to round a value stochastically: by comparing a random number to the residual and rounding up if it is smaller, or by adding a random number to the residual and letting the carry out from that control rounding.
Algorithms~\ref{alg:stochastic_rounding_by_comparison}~and~\ref{alg:stochastic_rounding_by_addition} demonstrate how to do both (in the algorithms $\ll$ and $\gg$ stand for binary shifts left and right).
Stochastic rounding by addition looks shorter, but both algorithms require 5 operations in the main rounding parts (saturation is the same in both cases).
Saturation logic is a standard check for overflow at both ends of the dynamic range and note that if the input and output numbers would be unsigned, only one comparison instead of two would be required.
It is also worth noting that RN mode can be implemented similarly to Algorithm~\ref{alg:stochastic_rounding_by_addition}, as shown in Algorithm~\ref{alg:rounding_to_nearest}.

By comparing Algorithms~\ref{alg:stochastic_rounding_by_addition}~and~\ref{alg:rounding_to_nearest}, notice that SR has an overhead of a PRNG plus one operation to mask off the top bits of the random number, compared with RN.

\section{Numerical experiments}

\renewcommand{\arraystretch}{1.15}
\begin{table*}[ht!]
	\centering
	\caption{Iterations until convergence of the harmonic series for different arithmetics. Sums and errors relative to binary64 result (double precision floating-point) at five millionth iteration. Floating-point data from \cite{hipr19}. Floating point formats are defined in \cite{ieee19}. Averaged sums are from running the experiment 50 times in s16.15 and s8.7 arithmetics with SR, each time with different PRNG seed. RD refers to round down mode.}
	\begin{tabular}{rlll} \toprule 
		Arithmetic & Sum at $i=5\times10^6$ & Error at $i=5\times10^6$ & Iterations to converge \\ \midrule
		binary64 & $16.002$ & $0$ & $2.81... \times 10^{14}$ \\
		binary32 & $15.404$ & $0.598$ & $2097152$ \\
		binary16 & $7.086$ & $8.916$ & $513$ \\ \midrule
		s16.15 RN & $11.938$ & $4.064$ & $65537$ \\
		s16.15 RD & $10.553$ & $5.449$ & $32769$ \\
		s8.7 RN & $6.414$ & $9.588$ & $257$ \\
		s8.7 RD & $5.039063$ & $10.963$ & $129$ \\ \midrule
		s16.15 SR & $\mathrm{Mean}=16.002 $& \multirow{2}{*}{$-0.000135765$} & \multirow{2}{*}{$2^{32}+1$} \\
		(50 runs) & $\mathrm{std.dev.}=0.012$ & & \\ \midrule
		s8.7 SR & $\mathrm{Mean}=11.205$ & \multirow{2}{*}{$4.797$} & \multirow{2}{*}{$2^{16}+1$} \\
		(50 runs) & $\mathrm{std.dev.}=0.242$ & & \\ \bottomrule
	\end{tabular}
	\label{table:harmonic-series0}
\end{table*}

The algorithm of choice for the proposed hardware accelerator is Algorithm~\ref{alg:stochastic_rounding_by_addition} and here we test it first in software simulation on an ARM968 processor using fixed-point arithmetic.
The main advantages of SR are in summation algorithms, with the data with rounding errors biased into one direction which dominates the final error in the result of the sum.
Following the approach taken by \cite{hipr19} we have applied the Algorithm~\ref{alg:stochastic_rounding_by_addition} of stochastic rounding in software on a basic \textit{recursive summation} algorithm evaluating the harmonic series; this series is a \textit{divergent series} but converges when implemented in limited precision arithmetic using recursive summation.
The series is defined as $\sum_{i=1}^{\infty}1/i=1+\frac{1}{2}+\frac{1}{3}+\cdots$ --- it can be seen that the addends are getting smaller while the total sum keeps increasing and as \cite{hipr19} reported the sum converges in floating-point arithmetic when the addends become small enough that they do not change the total sum anymore (due to very different exponents and round off error on addition).
This issue is called \textit{stagnation} in \cite{bhm20}, a problem which happens in summing algorithms in floating-point arithmetic.

This experiment was run in 32- and 16-bit fixed-point arithmetics.
The sum has a numerical type of s16.15 or s8.7 and is initialized to 1.
Then the series is started from $i=2$ and the division is done in either 32-bit or 16-bit fractional type (u0.32 or u0.16), followed by the addend being rounded to the sum's format with various rounding routines.
While fixed-point addition is known to be exact if there are no overflows, in this case it is not since the addends have more fractional precision than the sum.

Table~\ref{table:harmonic-series0} demonstrates the results with various fixed-point types; floating-point results are also provided for comparison.
Five million iterations were chosen to have a manageable run time, but the number of iterations to convergence is also reported.
As expected, most of the fixed-point types converge as soon as the addends in the series become small enough to be evaluated at lower than s16.15 precision, when the values cross $0.5\varepsilon$.
However, it can be seen that fixed point with SR can accurately replicate the sum of the binary64 format in 5 million iterations without converging.
Given that stochastic rounding is probabilistic rounding, it might
still produce some effect in later iterations stochastically and therefore it can be said that \textit{the harmonic sum with stochastic rounding never converges --- there is a diminishing, but non zero probability of rounding up the addends and affecting the sum}.
In practice it converges also when the numerical type of the addends runs out of bits and the probability of rounding up becomes 0.
This can also happen if there is a limited amount of random bits available for performing stochastic rounding and can especially be significant in rounding the double precision adder/accumulator results as these can be held in thousands of bits before rounding \cite{uguen:hal-01488916}.

This experiment provides a confirmation that Algorithm~\ref{alg:stochastic_rounding_by_addition} works as expected.
In summary, running the harmonic series 50 times in s16.15 with SR, it is shown that the averaged result has a very small error compared to the sum computed in binary64, while s16.15 with RN stagnates just after 65536 steps.

\section{Specification}
\label{sec:specification}

In this section we describe the specification of the proposed accelerator that was designed and included in the upcoming SpiNNaker2 neuromorphic chip \cite{hoppnerspinnaker2}.

The rounding and saturation accelerator is a memory mapped unit, connected through an AHB bus --- a set of memory addresses are allocated, for different rounding routines and numerical formats, to which arguments are written, and from which the rounded values are read out.
For rounding multiplication results, it is useful to have a $\text{64-bit} \rightarrow \text{32-bit}$ number rounding, with configurable rounding bit position from 0 to 31.
Given that the ARM M4F processor has 32-bit wide interfaces, two memory cycles are required for inputting 64-bit arguments through AHB into the accelerator.
For other use cases, $\text{32-bit} \rightarrow \text{32-bit}$, $\text{32-bit} \rightarrow \text{16-bit}$, and $\text{16-bit} \rightarrow \text{16-bit}$ round-and-saturate operations are also supported.
For these configurations, one memory cycle is required for input, therefore, aiming at single cycle for the main part of rounding, the accelerator will either have a 4- or a 3-cycle delay for a write-round-read operation for 64- or 32-bit arguments respectively.

Both signed and unsigned number types are supported, given a wide range of use cases for both types shown in \cite{doi:10.1098/rsta.2019.0052}.
Furthermore, as the ARM M4F has single precision floating-point hardware support, it might be beneficial to round binary32 to bfloat16 \cite{intel_bfloat16}.
This format can be useful for representing and storing neural network weights for example, which can be operated on using the floating-point unit by inputting into a higher part of the floating-point registers and then rounded back to bfloat16 before writing to memory.
Finally, given that an adder is required in stochastic rounding, we can also include RN mode (with rounding up on ties, for cheaper implementation), which can reuse the adder to add 0x1 shifted to the required rounding bit position.

\section{Design}
\label{sec:design}

\begin{figure}[t!]
	\centering
	\includegraphics[width=3.5in]{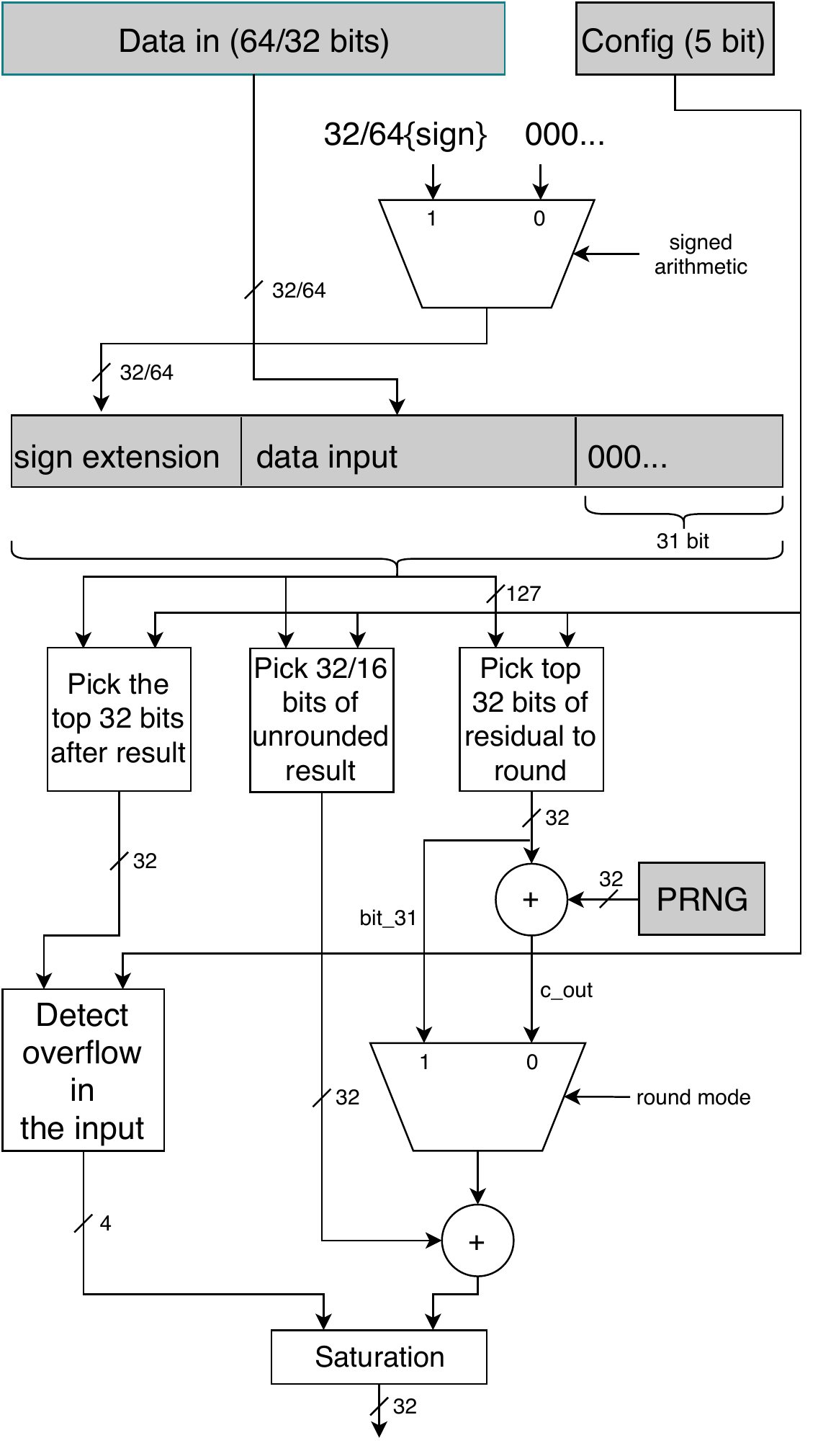}
	\caption{Architectural diagram for the rounding and saturation accelerator.}
	\label{fig:nmu_sr_architecture}
\end{figure}

Figure~\ref{fig:nmu_sr_architecture} gives an architectural diagram of the accelerator that performs rounding and saturation.
The functionality of the accelerator includes a combination of SR and RN modes, as in the Algorithms~\ref{alg:stochastic_rounding_by_addition}~and~\ref{alg:rounding_to_nearest}.
Signals \textit{signed arithmetic} and \textit{round mode} are derived from the address supplied by the AHB bus, depending on which address was written by the processor.

The main mechanism is to pick the top 32 bits of the residual depending on the configuration register, which is set up beforehand and contains the number of bits to round, 0 to 31 (0 means round 1 bit, 31 means round 32 bits).
Then, the 32 bits after these residual bits are also isolated which is an unrounded result at this point.
The minimum number of bits to round is 1, therefore the data input is extended to the right by 31 bits to support Verilog's \textit{base minus 32} bit slicing functionality.
For the same reason the input data is extended by 32 bits to the left, for overflow detection by operating on 32 bits after the result.

A pseudorandom number is added to the residual and the carry bit \textit{c\_out} is captured from that.
Then, depending on the \textit{round mode}, either the top bit of the residual (in case of RN mode) or the \textit{c\_out} (in case of SR mode) is added to the unrounded result which performs round-up if it is 1 and round-down if it is 0.
Finally the rounded result and the overflow bits are used to saturate the result if required.

\section{Evaluation}
\label{sec:evaluation}

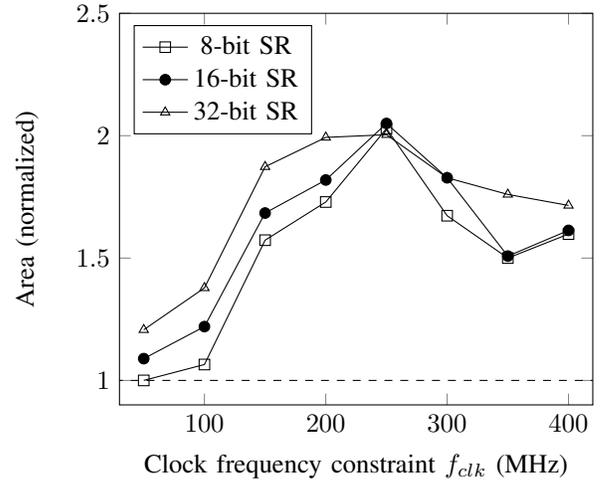
\begin{figure}[t!]
	\begin{center}
		\begin{tikzpicture}
		\begin{axis}[
		xlabel={Clock frequency constraint $f_{clk}$ (MHz)},
		ylabel={Area (normalized)},
        xmin=30, xmax=420,
        ymin=0.9, ymax=2.5,
		legend pos=north west,
		width=3.1in
		]
		
		\addplot[
		color=black,
		mark=square,
		]
		coordinates {
			(50, 1.0)(100, 1.065)(150, 1.573)(200, 1.729)(250, 2.029)(300, 1.673)(350, 1.5)(400, 1.598)
		};
		\addlegendentry{8-bit SR}
		
		\addplot[
		color=black,
		mark=*,
		]
		coordinates {
			(50, 1.089)(100, 1.22)(150, 1.684)(200, 1.819)(250, 2.05)(300,1.828)(350,1.508)(400,1.613)
		};
		\addlegendentry{16-bit SR}
		
		\addplot[
		color=black,
		mark=triangle,
		]
		coordinates {
			(50, 1.207)(100, 1.378)(150, 1.873)(200, 1.993)(250, 2.004)(300, 1.828)(350,1.76)(400,1.715)
		};
		\addlegendentry{32-bit SR}
		
		\addplot[
		color=black,
		dashed,
		]
		coordinates {
			(0, 1)(1000, 1)
		};
		
		\end{axis}
		\end{tikzpicture}
	\end{center}
	\caption[Circuit area of the stochastic rounding accelerator]{Circuit area of the accelerator when synthesized with different clock constraints.
	Three accelerator versions are shown with 8-, 16-, and 32-bit stochastic rounding.}
	\label{fig:area-comparison-sr}
\end{figure}

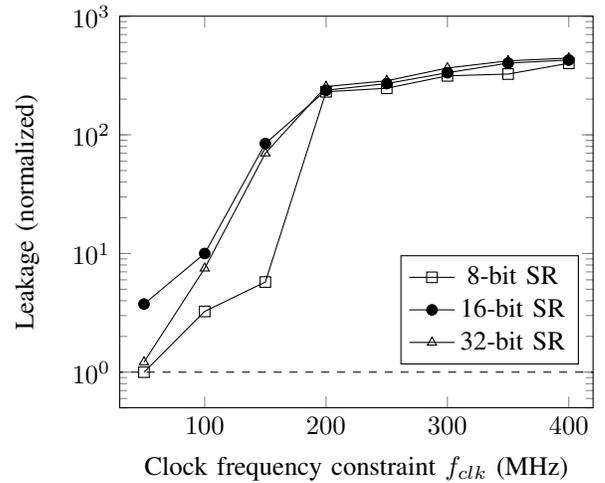
\begin{figure}[t!]
	\begin{center}
		\begin{tikzpicture}
		\begin{axis}[
		ymode=log,
		xlabel={Clock frequency constraint $f_{clk}$ (MHz)},
		ylabel={Leakage (normalized)},
         xmin=30, xmax=420,
        ymin=0, ymax=1000,
        legend style = {at = {(0.97, 0.25)}, anchor = east},
		width=3.1in
		]
		
		\addplot[
		color=black,
		mark=square,
		]
		coordinates {
			(50, 1.0)(100, 3.25)(150, 5.75)(200, 230.75)(250, 247)(300, 314.25)(350, 325.25)(400, 401)
		};
		\addlegendentry{8-bit SR}
		
		\addplot[
		color=black,
		mark=*,
		]
		coordinates {
			(50, 3.75)(100, 10)(150, 84.5)(200, 237.5)(250, 270.5)(300,333.75)(350,402.25)(400,427)
		};
		\addlegendentry{16-bit SR}
		
		\addplot[
		color=black,
		mark=triangle,
		]
		coordinates {
			(50, 1.216)(100, 7.5)(150, 70)(200, 255.25)(250, 285.5)(300, 366)(350,420.25)(400, 443.5)
		};
		\addlegendentry{32-bit SR}
		
		\addplot[
		color=black,
		dashed,
		]
		coordinates {
			(0, 1)(1000, 1)
		};
		
		\end{axis}
		\end{tikzpicture}
	\end{center}
	\caption[Leakage of the stochastic rounding accelerator]{Leakage of the accelerator when synthesized with different clock constraints.
		Three accelerator versions are shown with 8-, 16-, and 32-bit stochastic rounding.}
	\label{fig:leakage-comparison-sr}
\end{figure}

The main logical path of the accelerator contains two adders --- one 32-bit for rounding and one 8-, 16-, or 32-bit for the stochastic rounding part when a random number is added to the residual.
The architectural diagram in Figure~\ref{fig:nmu_sr_architecture} demonstrates a 32-bit version, but it is worth evaluating the three versions as there is some evidence that not all of the 32 bits are needed in SR, as shown in the previous work \cite[Sec.~5c(iii)]{doi:10.1098/rsta.2019.0052}.
All of the logic, except some saturation checks, are performed in a single cycle.
Saturation logic contains basic checks of the overflow flags depending on the address input from the AHB bus and in our implementation is done on the AHB output cycle.

Following the synthesis study approach taken by us before \cite{8464785}, we have executed it on the current accelerator using the makeChip hosted design service platform \cite{makechip-online} for the GLOBALFOUNDRIES 22FDX technology \cite{7838029} for which the SpiNNaker2 chip is being developed.
An ultra-low voltage \textit{8t-CNRX} standard-cell library with multiple voltage threshold options is used for implementation.
The standard cells use the \textit{adaptive body biasing} (ABB) technique for post-silicon adaptation of transistor threshold voltage \cite{8901768, 8932444}.
Namely, two main categories of cells are used: Low-Voltage-Threshold (further called LVT) and Super-Low-Voltage-Threshold (further called SLVT) cells --- the former with the larger propagation delay but significantly less leakage than the latter, much faster, cells.
A nominal supply voltage of \SI{0.5}{\volt} is considered for low power operation.
Due to manufacturing variations, synthesis is performed in a worst case speed condition at \SI{0.45}{\volt} and \SI{-40}{\celsius}.
Three versions of the accelerator are synthesized varying the clock frequency constraint
\begin{equation*}
f_{clk}=\{50, 100, 150, 200, 250, 300, 350, 400\} \: \text{MHz},
\end{equation*}
and the leakage power as well as area is measured.

Figure~\ref{fig:area-comparison-sr} shows the area comparison of the three accelerators for different clock constraints and Figure~\ref{fig:leakage-comparison-sr} shows leakage.
From this data it can be seen that at low clock frequencies, the adder width can save some area and leakage (more than an order of magnitude less leakage with 8-bit SR at $f=\SI{150}{\mega\hertz}$), but at higher frequencies other costs dominate and the savings are not that evident anymore.
Especially for leakage; the leakage of the circuit apart from the adder dominates the total and changing to a smaller adder does not produce significant improvements.
Notice that the circuit area is largest at the intermediate frequency of $f_{clk}=250$~MHz, which most likely can be explained by LVT cells being replaced with SLVT cells (smaller cells or lower number of cells) on the critical path in $f_{clk} > 250$~MHz settings, although a more thorough investigation of synthesis would be required to explain this.

\begin{figure}[htbp]
	\centering
	\includegraphics[width=3in]{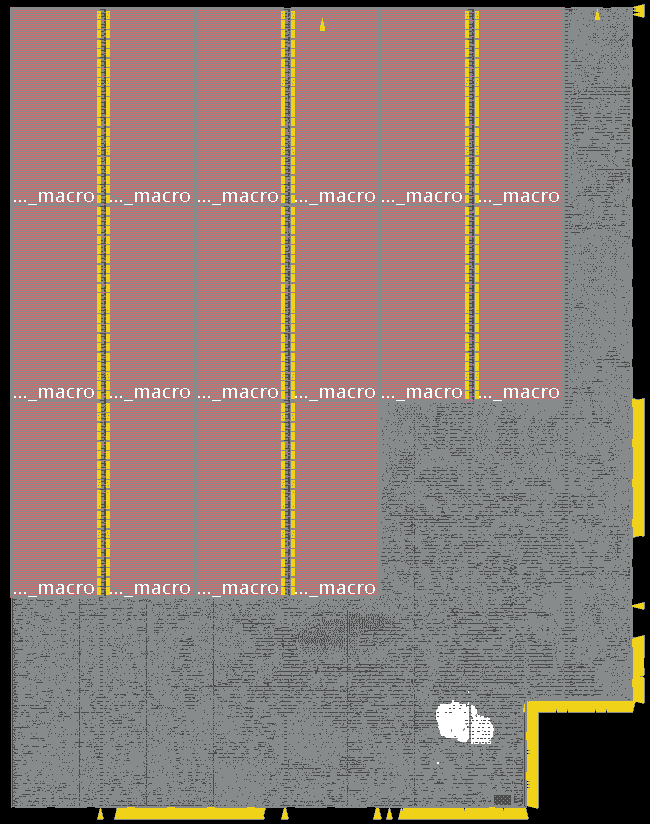}
	\caption[Layout of the processing element (PE) after place and route]{Layout of a processing element (PE) after place and route. Cells marked \textit{...macro} bundled at north-west corner belong to local SRAM. The rest of the cells at the south-east corner belong to an ARM M4F based PE.
		Out of that, cells highlighted in white belong to the rounding accelerator (picture provided by Stefan Scholze).}
	\label{fig:layout_sr}
\end{figure}

Figure~\ref{fig:layout_sr} shows the rounding accelerator highlighted in a layout of a single PE.
The area of the accelerator is estimated at \SI{1004}{\micro \meter \squared}.

\section{Conclusion}

We have presented algorithms and an accelerator for performing rounding and saturation of numbers up to 64 bits, including stochastic rounding which is becoming popular in machine learning.
This includes rounding of fixed-point/integer values at any bit position as well as binary32 to bfloat16 rounding.
The chosen SR algorithm was tested on the harmonic series computed with a basic recursive summation, demonstrating how SR can help avoid numerical \textit{stagnation} in fixed-point arithmetic.
Evaluation of the accelerators with different precisions of SR step was performed, showing an order of magnitude of leakage improvement with 8-bit SR at $f=\SI{150}{\mega\hertz}$.
The accelerator will be included in the SpiNNaker2 chip, which is scheduled for 2020 release and is based on an ARM Cortex-M4F processor.
Since this processor does not provide a wide array of rounding and saturation instructions to support fast fixed-point arithmetic, especially mixed-format fixed-point arithmetic, this accelerator will complement it and provide that functionality.

The presented results should also be applicable in implementing stochastic rounding of floating-point arithmetic, such as rounding the extended precision results from the floating-point adder or multiplier \cite{ibm-sr-patent-add, ibm-sr-patent-mult}.

\section{Acknowledgements}

The author thanks to Sebastian H\"{o}ppner, Stefan Scholze, and Andreas Dixius of Technical University of Dresden for the help with the synthesis tools, as well as Nicholas J. Higham for his comments on the manuscript.
The work was funded by the Kilburn studentship and an EPSRC Doctoral Prize Fellowship.

\bibliographystyle{IEEEtran}
\IEEEtriggeratref{14}
\bibliography{IEEEabrv,references,../njhigham-bib/strings,../njhigham-bib/njhigham}

\end{document}